\documentclass[sigconf]{acmart}

\usepackage{enumerate}
\usepackage{enumitem}
\usepackage{multirow}
\usepackage{etoolbox}
\usepackage{fontawesome}
\usepackage{array}
\usepackage{cleveref}
\usepackage{subcaption}
\usepackage{tcolorbox}
\usepackage{caption}
\usepackage{color, colortbl}
\definecolor{Gray}{gray}{0.9}
\definecolor{bg}{RGB}{255,249,227}
\usepackage[font=itshape,begintext=``,endtext='']{quoting}

\newcolumntype{P}[1]{>{\raggedright\arraybackslash}p{#1}}
\newcolumntype{C}[1]{>{\centering\arraybackslash}p{#1}}

\acmConference[ESEM '21]{ESEM '21: ACM/IEEE International Symposium on Empirical Software Engineering and Measurement}{October 11--15, 2021}{Bari, Italy}
\acmISBN{978-1-4503-XXXX-X/18/06}
\acmSubmissionID{141}
\setcopyright{acmcopyright}

\begin{document}
	
\title{Towards a Human Values Dashboard for Software Development: An Exploratory Study}

\author{Arif Nurwidyantoro}
\affiliation{%
	\institution{Monash University}
	\city{Melbourne}
	\country{Australia}}
\email{arif.nurwidyantoro@monash.edu}
\author{Mojtaba Shahin}
\affiliation{%
	\institution{Monash University}
	\city{Melbourne}
	\country{Australia}}
\email{mojtaba.shahin@monash.edu}
\author{Michel Chaudron}
\affiliation{%
	\institution{Eindhoven University of Technology}
	\city{Eindhoven}
	\country{Netherlands}}
\email{m.r.v.chaudron@tue.nl}
\author{Waqar Hussain}
\affiliation{%
	\institution{Monash University}
	\city{Melbourne}
	\country{Australia}}
\email{waqar.hussain@monash.edu}
\author{Harsha Perera}
\affiliation{%
	\institution{Monash University}
	\city{Melbourne}
	\country{Australia}}
\email{harsha.perera@monash.edu}
\author{Rifat Ara Shams}
\affiliation{%
	\institution{Monash University}
	\city{Melbourne}
	\country{Australia}}
\email{rifat.shams@monash.edu}
\author{Jon Whittle}
\affiliation{%
	\institution{CSIRO's Data61}
	\city{Melbourne}
	\country{Australia}}
\email{jon.whittle@data61.csiro.au}


\newtoggle{draft}
\toggletrue{draft}
\iftoggle{draft}{%
	\newcommand{\idea}[1]{{ \color{black}{\textit{#1}}\xspace}}
	\newcommand{\todo}[1]{{\color{red}{TODO:\textbf{#1}}\xspace}}
	\newcommand{\jon}[1]{{\color{red}{Jon: #1}\xspace}}
	\newcommand{\waqar}[1]{{\color{blue}{Waqar: #1}\xspace}}
	\newcommand{\rifat}[1]{{\color{violet}{Rifat: #1}\xspace}}
	\newcommand{\harsha}[1]{{\color{purple}{Harsha: #1}\xspace}}
	\newcommand{\arif}[1]{{\color{violet}{#1}\xspace}}
	\newcommand{\Mojtaba}[1]{{\color{blue}{#1}\xspace}}
	\newcommand{\michel}[1]{{\color{olive}{Michel: #1}\xspace}}
	\newcommand{\newtext}[1]{{\color{blue}{#1}}}
}{%
	\newcommand{\idea}[1]{}
	\newcommand{\todo}[1]{}
	\newcommand{\jon}[1]{}
	\newcommand{\waqar}[1]{}
	\newcommand{\rifat}[1]{}
	\newcommand{\harsha}[1]{}
	\newcommand{\arif}[1]{#1}
	\newcommand{\Mojtaba}[1]{}
	\newcommand{\michel}[1]{}
	\newcommand{\newtext}[1]{#1}
}%
\newcommand{\say}[3]{
    \begin{itemize}[leftmargin=1.5em]
        \item[\faCommentingO] ``\textit{#1}'' (#2 - #3)
    \end{itemize}
}

\newcommand{\sayt}[1]{
``\textit{#1}''
}
\newcommand{\sr}{\rule[-0.45cm]{0pt}{0.5cm}}
\newcommand{\srb}{\rule[-0.45cm]{0pt}{1mm}}
\newcommand{\summary}[1]{
\begin{center}
\begin{tcolorbox}[colback=black!5!white,colframe=black!75!black]
\textit{#1}
\end{tcolorbox}
\end{center}
}

\begin{abstract}
\textbf{Background}: There is a growing awareness of the importance of human values (e.g., inclusiveness, privacy) in software systems. However, there are no practical tools to support the integration of human values during software development. 
We argue that a tool that can identify human values from software development artefacts and present them to varying software development roles can (partially) address this gap. We refer to such a tool as \textit{human values dashboard}.
Further to this, our understanding of such a tool is limited. 
\textbf{Aims}: This study aims to (1) investigate the possibility of using a human values dashboard to help address human values during software development, (2) identify possible benefits of using a human values dashboard, and (3) elicit practitioners' needs from a human values dashboard.
\textbf{Method}: We conducted an exploratory study by interviewing 15 software practitioners. A dashboard prototype was developed to support the interview process. We applied thematic analysis to analyse the collected data.
\textbf{Results}: Our study finds that a human values dashboard would be useful for the development team (e.g., project manager, developer, tester). Our participants acknowledge that development artefacts, especially requirements documents and issue discussions, are the most suitable source for identifying values for the dashboard. Our study also yields a set of high-level user requirements for a human values dashboard (e.g., it shall allow determining values priority of a project). 
\textbf{Conclusions}: Our study suggests that a values dashboard is potentially used to raise awareness of values and support values-based decision-making in software development. Future work will focus on addressing the requirements and using issue discussions as potential artefacts for the dashboard.
\end{abstract}

\ccsdesc[500]{Software and its engineering~Software creation and management}

\keywords{Human Values, Dashboard, Software Development}

\maketitle

\section{Introduction}

Human values such as inclusiveness, social justice, and privacy, or \textit{`what people hold important in their life'} \cite{Schwartz2012,rokeach1973nature}, have received increased attention in the last couple of years in the software industry. A number of recent events have indicated that people's awareness has grown about human values, and they strongly react to the violation of their values in software.
As an example, changes in WhatsApp's terms and privacy policy led millions of its users to migrate to alternative messaging apps~\cite{best2021}. One of the reasons surfaced was the loss of trust in Facebook that will have access to WhatsApp's users' data after the new policy comes into play~\cite{best2021}. In this event, users seem to remember the infamous Facebook's Cambridge Analytica privacy case a few years back~\cite{Confessore2018}. Another example is that digital and human rights groups protested the use of facial recognition systems in justice systems that introduce bias against minorities~\cite{Schapiro2020}. The bias caused minorities more likely to be detected as offenders and increase the fear of unfair treatments~\cite{Schapiro2020}. These events are aligned with a characteristic of human values where people feel threatened when their values are jeopardised~\cite{Schwartz2012}. To prevent these situations, consideration of values is necessary because it could influence acceptance from users towards an application \cite{Wang2013,Harris2016,Fu2013}.

Addressing human values is difficult because of their subjective nature \cite{Winter2018} and their definition depends on the context they are applied to \cite{Kujala2009,Mougouei2018}. Several solutions have been proposed to support practitioners address values in software. These solutions are commonly in the form of frameworks, techniques, practices, and guidelines, such as Value-based Requirements Engineering~\cite{Thew2018}, Value Sensitive Design~\cite{Friedman2008,Friedman2013}, or Continual Value(s) Assessment~\cite{PereraRE2020}.
However, these solutions aim to consider values at a specific phase of software development, such as requirements or design phase, or satisfy a specific type of practitioner (e.g., designer). Furthermore, these solutions are not necessarily equipped with a tool to facilitate the understanding and integration of values during software development. We claim that providing a \textit{human values dashboard} can bridge this gap and help practitioners effectively and efficiently understand and handle values during software development.

In software development, dashboards are commonly used to support decision-making ~\cite{Ivanov2018,Ivanov2018a}, promote awareness within a project~\cite{Treude2010,Baysal2013}, and monitor development activities~\cite{Leite2015}.
It is common for a dashboard in software development to use development artefacts as its source. For instance, Leite et al. developed a dashboard that used commit history to detect unusual events~\cite{Leite2015}. Several other dashboards have also been developed using artefacts from software repositories~\cite{github-personal-dashboard,github-organizational-dashboard,cauldron,mautic}. 
In terms of values in software development artefacts, previous studies have investigated a few values, such as security~\cite{Fischer2017,Viega2002,Pletea2014,Alqahtani2017}, privacy~\cite{Kim2012,Li2015,Gibler2012,Naseri2019,Kuznetsov2016,Slavin2016,Sharma2014}, or energy efficiency~\cite{Bao2016,Pereira2017}. Although these works not specifically addressed security, privacy, or energy efficiency as values, the works show the possibility to discover values in the artefacts.
A dashboard is suitable for our purpose because it allows information to be displayed visually to \textit{`facilitate understanding'}~\cite{Wexler2017}.
We believe a dashboard can help clarify the less-known and abstract concept of values~\cite{Mougouei2018,Perera2020} to software practitioners.

\begin{figure}[bth]
  \centering
  \includegraphics[width=\linewidth]{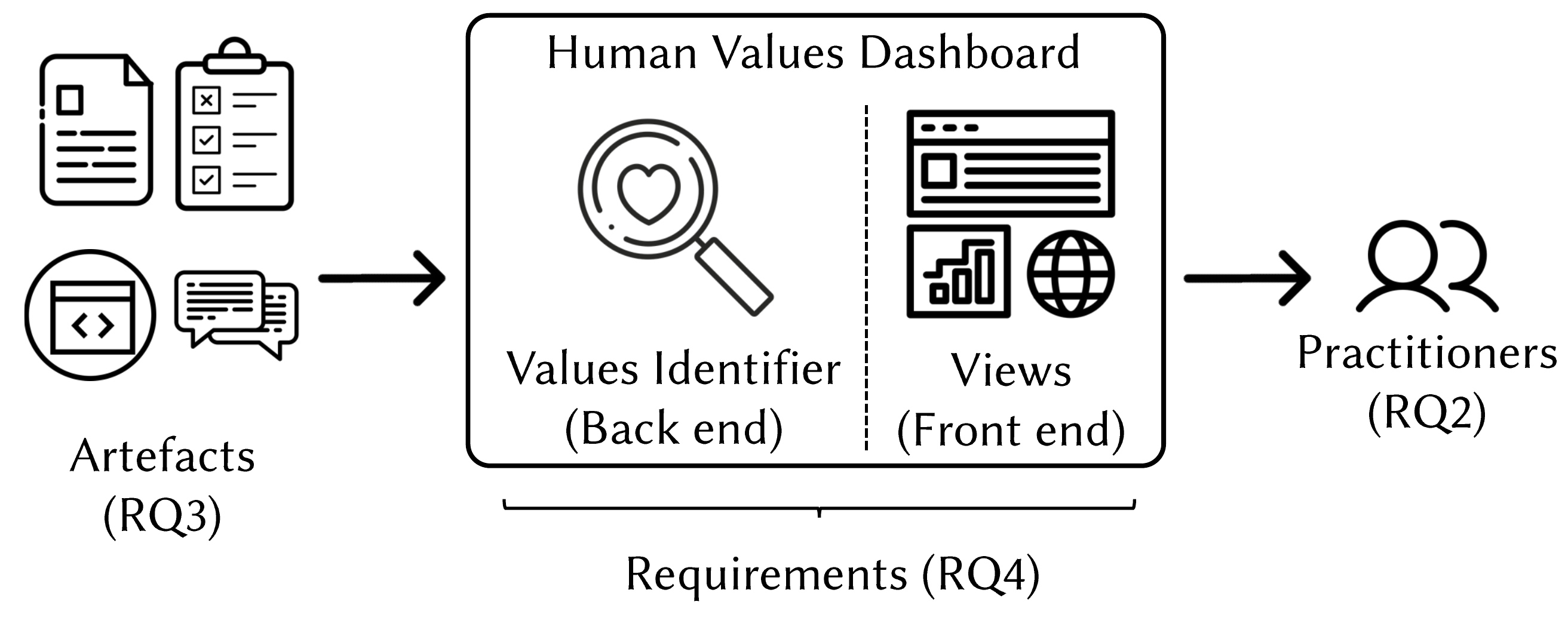}
  \caption{Human values dashboard and research questions}
  \label{fig:proposed-dashboard}
  \vspace{-3mm}
\end{figure}

\begin{figure}[bth]
  \vspace{-3mm}
  \centering
  \includegraphics[width=\linewidth]{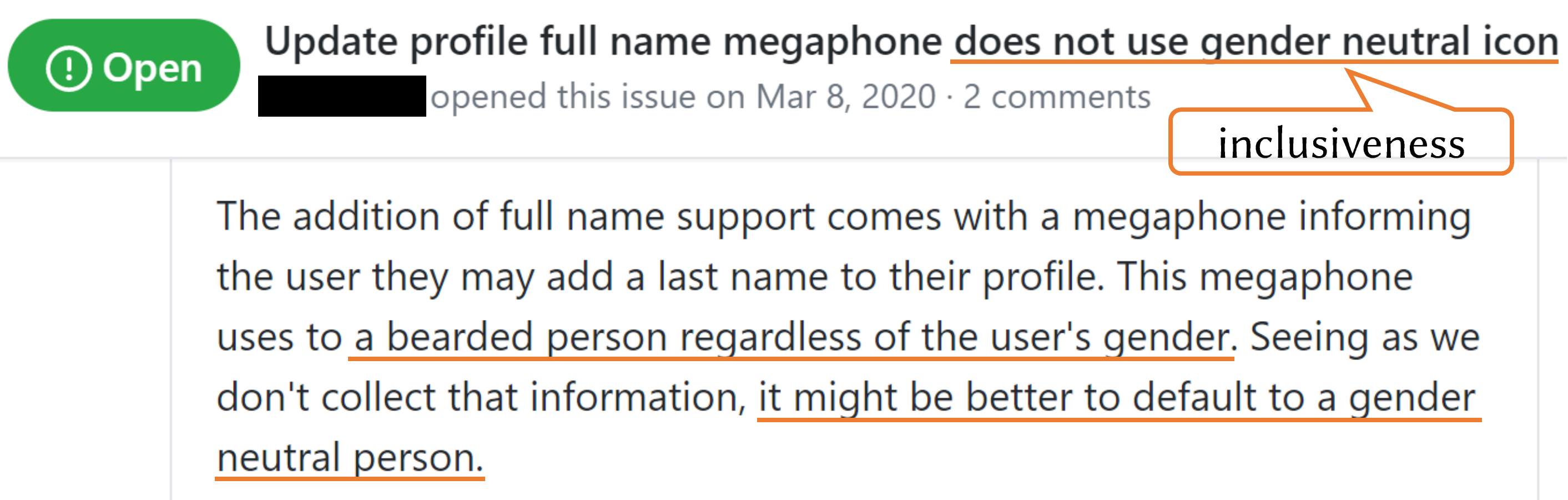}
  \caption{Example of values identified in an issue discussion}
  \label{fig:values-issues}
  \vspace{-3mm}
\end{figure}

\Cref{fig:proposed-dashboard} presents our vision of a human values dashboard that uses software development artefacts as its data source and displays values identified in the artefacts to support practitioners addressing those values in the software. 
To this end, we propose a human values dashboard consisting of a back end and a front end. The back end of the dashboard provides functionality to identify values from software development artefacts. The identification of values could be made manually (e.g., by the development team) or using an automated approach. The back end is necessary because these artefacts naturally do not have values identified in them yet. For example, \Cref{fig:values-issues} shows a user of an open source application expresses his/her opinion of \textit{inclusiveness} to be present in the application in an issue discussion. Based on this example, we define a human value can be identified in a software development artefact if \textit{there is a notion of that value in the artefact}. The front end of the dashboard displays values identified from various artefacts in different views (for different roles). 
In this paper, we used the terms \textit{human values dashboard} and \textit{values dashboard} interchangeably.

Before developing such values dashboard, we need to understand what is necessary for a human values dashboard to be helpful in software development. This study serves as an initial stage to inform our objective to develop a human values dashboard by focusing on the following research questions: 
\begin{enumerate}[label=\textbf{RQ\arabic*}]
    \item What are the perceptions of practitioners towards human values in software development?
    
    \item Who will benefit, and what is the benefit of a human values dashboard? 
    
    \item Are software development artefacts suitable for identifying values for the dashboard? If so, which artefacts?
    
    \item What is needed for a human values dashboard to be helpful in software development?
\end{enumerate}
\Cref{fig:proposed-dashboard} shows how the last three RQs relate to the visioned dashboard.

To answer those research questions, we conducted an exploratory study. We started with developing a human values dashboard prototype. Then, we demonstrated the prototype and interviewed 15 software practitioners to obtain their insights.

Our results reveal that our participants agree that human values are important in software, such that a supporting tool will be beneficial. The participants also believe that a values dashboard can benefit various software development roles (e.g., \textit{project manager}, \textit{developer}, \textit{tester}), primarily to raise awareness of values and support values-based decision-making in project management. Our practitioners also acknowledge that software development artefacts are suitable as the source for the dashboard. Among those artefacts, requirements documents and issue discussions are deemed as the most suitable. Our study suggests a set of high-level requirements to inform the future development of a human values dashboard (e.g., \textit{it shall display the artefacts based on the values priority determined in a project}).

The remaining sections of this paper are organised as follows. \Cref{sec:backgrounds} summarises previous studies that are relevant to this study. In \Cref{sec:methodology}, we describe our methodology. We report our findings in \Cref{sec:result}. \Cref{sec:discussion} discusses the findings and proposes future directions. \Cref{sec:threats} describes the threats to the validity of our study. Finally, \Cref{sec:conclusion} concludes the paper.

\section{Backgrounds and Related Work}
\label{sec:backgrounds}

\subsection{Human Values}
\subsubsection{Human values definition}
Human values, such as \textit{achievement} or \textit{benevolence}, are defined by Schwartz as `\textit{things that people hold important in their life}' \cite{Schwartz2012}. Meanwhile, Rokeach defined values as `\textit{a belief that a particular way of doing something is personally or socially preferable to the opposite ways}'~\cite{rokeach1973nature}. 
Studies in social sciences suggested a degree of relative importance between these values for each person~\cite{rokeach1973nature,Schwartz2012,Schwartz2017}. Because values are \textit{`intertwined with feelings'}~\cite{Schwartz2012}, a threat to someone's values can upset that person. Otherwise, consideration of values will bring enjoyment for a person. For example, in a software engineering context, a user of an application who values inclusiveness may expect this value to be present in the application (\Cref{fig:values-issues}). 

\begin{figure}[bth]
  \centering
  \includegraphics[width=\linewidth]{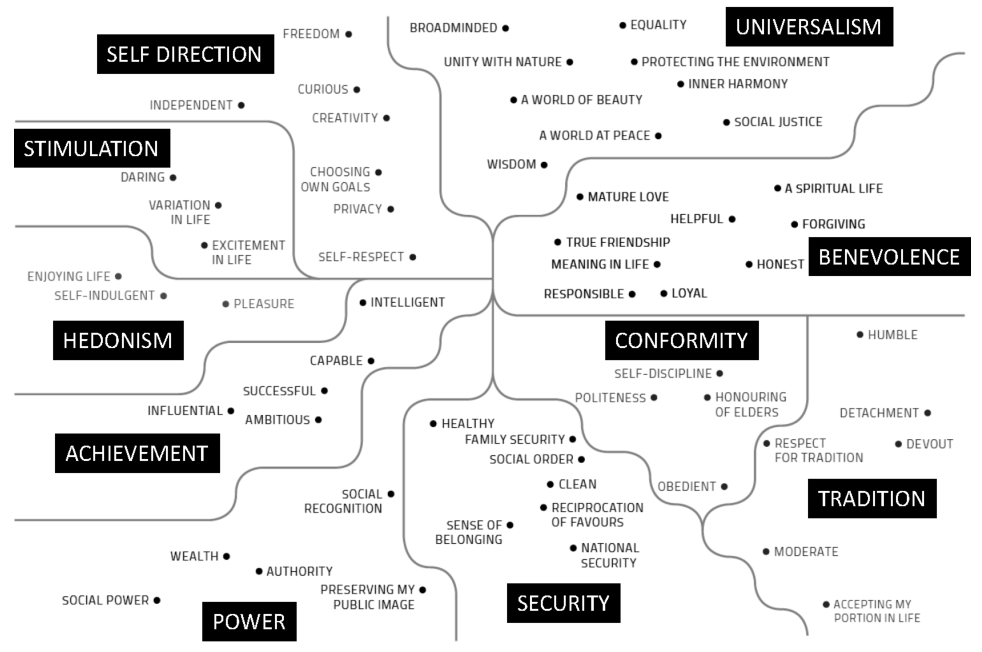}
  \caption{The Schwartz models of basic human values \cite{Schwartz2012} taken from \cite{holmes_blackmore_hawkins_wakeford_2011}}
  \label{fig:values-model}
  \vspace{-3mm}
\end{figure}

Social sciences have proposed several models that identified human values and divided them into several categories~\cite{rokeach1973nature,Bird1998,Cheng2010,Schwartz2012,Schwartz2017}. Among those models, Schwartz's model~\cite{Schwartz2012} is considered the most complete as it covers the largest number of values compared to other models \cite{Cheng2010}. 
In this study, we used Schwartz's model (\Cref{fig:values-model}) to introduce values to software practitioners during data collection. 

\subsubsection{Human Values and Non-Functional Requirements}
Some values, such as \textit{privacy} and \textit{security}, may have a similar concept in non-functional requirements (NFRs). In software development, NFRs are associated with quality properties that can influence the degree of satisfaction of a software~\cite{Mairiza2010, Glinz2007,Barn2016}. 
According to Barn~\cite{Barn2016}, human values are distinct from NFRs, but they can benefit from utilising existing NFRs frameworks. The rationale was coming from an argument that `\textit{human values are associated with the moral concerns of users instead of using system perspective in NFRs}'~\cite{Barn2016}. We support this idea with an argument that human values have a much broader sense than NFRs by including non-technical aspects important for users. 

\subsection{Human Values in SE}
\subsubsection{Solutions to integrate values in software}
Several solutions have been proposed to integrate human values into software engineering. For example, Value-based Requirement Engineering (VBRE) was introduced for the requirements engineering stage to elicit values from users and stakeholders~\cite{Thew2018}. Another approach called Value Sensitive Design (VSD)~\cite{Friedman2008,Friedman2013} was proposed to integrate the consideration of values into the design process of a system. A framework called Continual Value(s) Assessment (CVA) was also proposed to extend a set of an application's functionalities based on an evaluation of value implications of the existing functionalities~\cite{PereraRE2020}. 

These solutions proposed values to be considered in specific stages of software development, especially in the early stages, such as requirements and design. We believe it is possible to support the integration of values in the later stages of the development (e.g., implementation). For instance, Hussain et al. have identified several places to introduce values in the whole software development phases in the SAFe Agile framework~\cite{hussain2021human}. 
In addition, these works proposed solutions as methods or frameworks. We argued that to be practical, a solution could also be in the form of a tool. Meanwhile, not so many studies proposed a tool to support values in software. Our study addressed this gap by visioning a dashboard as a solution. 
We believe our idea of a human values dashboard has the potential to support various stages of software development by utilising artefacts generated during software development as its data source.

\subsubsection{Human values in software development artefacts}
Software development activities normally generate artefacts. For instance, requirements documents are written as a result of requirements gathering activities. The development team may also discuss an issue report within repositories. These artefacts have been used by previous works to investigate human values. These studies mostly considered more familiar values in software engineering, such as \textit{security}, \textit{privacy}, and \textit{energy efficiency}. For example, some studies have investigated the notion of security in source codes~\cite{Fischer2017,Viega2002} and issue discussions~\cite{Pletea2014,Alqahtani2017}. 
Privacy has received a lot of attention by several investigations on various artefacts, such as source code and configuration files~\cite{Kim2012,Li2015,Gibler2012,Naseri2019,Kuznetsov2016}, application programming interfaces (API)~\cite{Slavin2016}, and project documentation~\cite{Sharma2014}. 
Other studies focused on the energy efficiency of an application~\cite{Bao2016,Pereira2017}. Although these solutions are related, they do not specifically consider security, privacy, or energy efficiency as values. However, these studies support our idea that values are present in software development artefacts. Hence, development artefacts can be suitable as the data source of a values dashboard.

\subsection{Dashboard for Software Development}
A dashboard is generally used in organisations to monitor progress \cite{Wexler2017} and support decision-making \cite{Janes2013}. In software development, recent studies had demonstrated the use of a dashboard to make decisions~\cite{Ivanov2018,Ivanov2018a} and promote awareness about the software project to the development team~\cite{Treude2010,Baysal2013}. For instance, Leite et al. proposed a dashboard to make developers aware of unusual events in repositories~\cite{Leite2015}. Another study used a dashboard to visualise concerns in the context of software evolution~\cite{Treude2009}. In practice, software projects use dashboards during development to monitor the development activities of a project~\cite{cauldron,mautic,github-personal-dashboard,github-organizational-dashboard}. 
In this study, we aimed to utilise those benefits of a dashboard, especially to promote the awareness of values during software development to practitioners.
\section{Methodology}
\label{sec:methodology}

We used semi-structured interviews supported with prototyping to answer our research questions. Prototyping is commonly used for requirements elicitation to \textit{`provide users with an idea of how a system will behave'} \cite{Fernandes2016}. 
Developing an artefact as a prototype to accompany an interview is also commonly used in the field of information systems \cite{Gasparic2017} and considered as the first iteration in the design science research methodology \cite{Pestana2018,Pestana2020,Bunde2021}.
Prototyping helped us communicate our ideas and obtain feedback from the interview participants to develop a human values dashboard.
In this study, we first developed a prototype of our visioned human values dashboard. Then, we conducted interviews with software practitioners. Finally, we analysed the interviews to address the research questions. Note that we obtained ethics approval from our university for this study. We also obtained written informed consent from the participant.

\begin{figure*}[htb]
  \centering
    \begin{subfigure}[b]{0.98\linewidth}
        \centering
        \includegraphics[width=\linewidth]{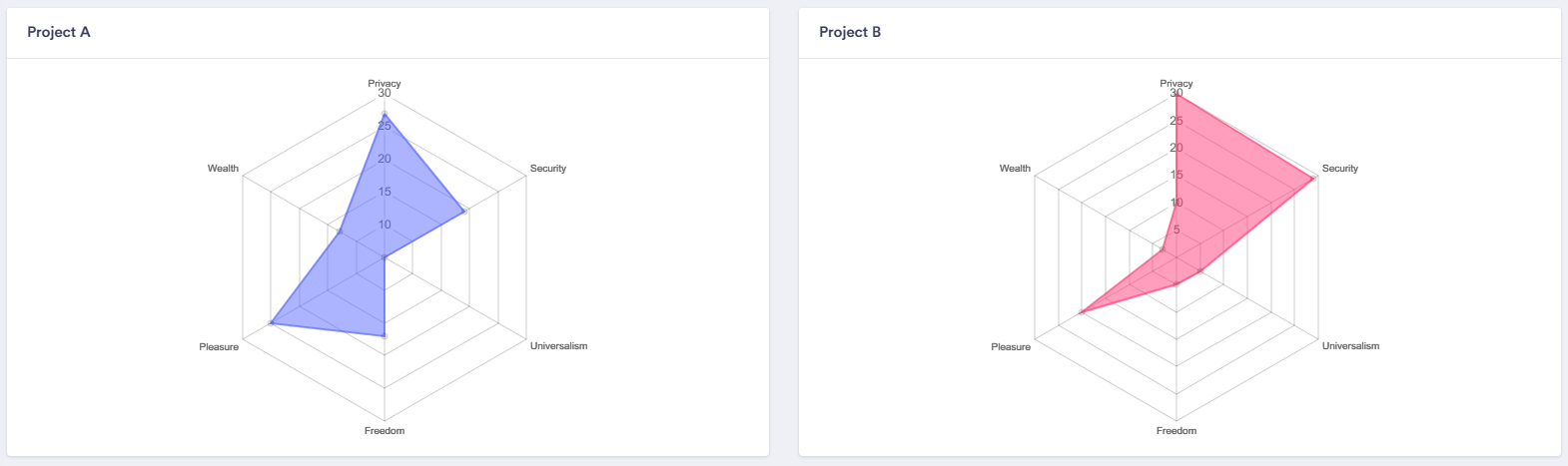}
        \caption{Summarised values overview (OV)}
        \label{fig:prototype-overview}
    \end{subfigure}
    \par\medskip
    \begin{subfigure}[b]{0.48\linewidth}
        \centering
        \includegraphics[width=\linewidth]{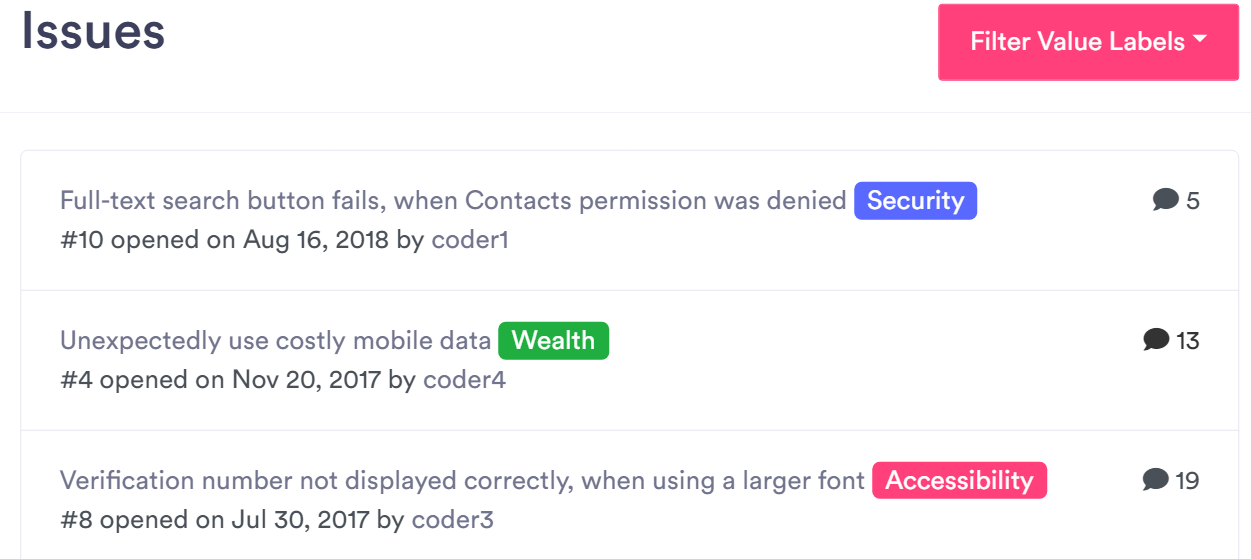}
        \caption{Values-labelled list (LI)}
        \label{fig:prototype-list}
    \end{subfigure}
    \hfil
    \begin{subfigure}[b]{0.48\linewidth}
        \centering
        \includegraphics[width=\linewidth]{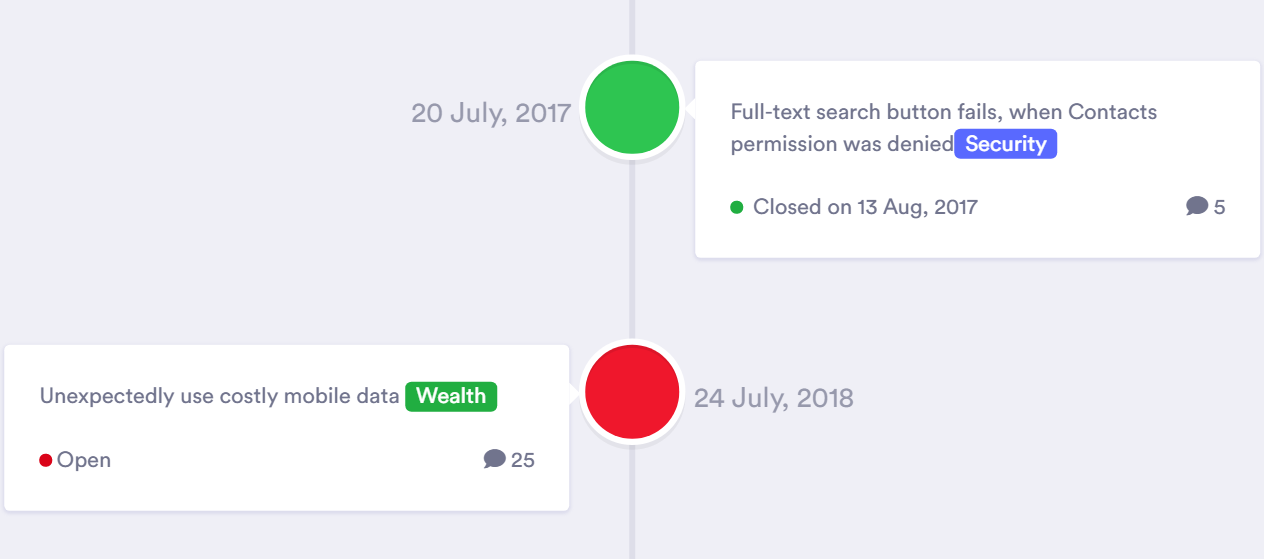}
        \caption{Values-labelled timeline (TM)}
        \label{fig:prototype-timeline}
    \end{subfigure}
    \caption{Dashboard View prototype}
    \label{fig:prototype}
    \vspace{-3mm}
\end{figure*} 

\vspace{-2mm}
\subsection{Prototype Development}
\label{subsec:dashboard}
We developed a prototype that displays values identified in software development artefacts to communicate our idea of a human values dashboard.  
The development of the dashboard prototype started with examining the literature, especially the studies related to dashboard development (e.g., \cite{Treude2009,Treude2010,Janes2013,Ivanov2018,Ivanov2019}).
We also looked at the existing dashboards for software development~\cite{github-personal-dashboard, github-organizational-dashboard, cauldron, mautic} to discover how they display information to support software development. These studies  
resulted in the following:
\begin{enumerate}[leftmargin=0.7cm,label=\textbf{(F\arabic*)}]
    \item Existing dashboards for software development have a common main objective to monitor the software development process, such as how many activities are happening or remaining issues to be addressed. This functionality allows the development team to be aware of the status of the project. We presumed displaying values in a dashboard could be beneficial to promote awareness for values to the development team. 
    \item Software practitioners prefer to use a dashboard to support their operational tasks~\cite{Ivanov2018}. Because these tasks could be different for each role, we thought it is necessary to develop several dashboard views to cater to various roles in software development.
    \item Artefacts from software repositories could be used as the data source for a dashboard \cite{cauldron,mautic}. Regarding this, we believed the vast amount of data in repositories is suitable for a dashboard. Furthermore, practitioners are accustomed to repositories in their daily activities.
\end{enumerate}

We then used those findings to guide our decisions in developing a prototype of human values dashboard. As a starting point, we chose issues (e.g., GitHub Issues) as the source for values identification to be displayed in the dashboard prototype (\textbf{F3}). This was because issues represent tasks that need to be addressed by the development team \cite{github-issues,gitlab-issues}. Furthermore, this is a place where a discussion about values may happen (\Cref{fig:values-issues}). We wanted the prototype to display issues and their identified values to support practitioners in addressing those values during development activities (\textbf{F1} and \textbf{F2}).
Afterwards, in the dashboard, we identified and displayed which values, in which issues they are found, and the number of issues where values are discovered in the software projects. 
We hypothesised that these measures could cater to various software development roles (\textbf{F2}). For instance, information on which values and in which issues values are identified will help developers and testers. Meanwhile, the number of issues where values are discovered could be useful for a project manager to ensure those issues are addressed properly. We argued that these measures could support practitioners in addressing those values during development activities (\textbf{F1} and \textbf{F2}).

Afterwards, we sampled and labelled a few issues from a random open source project repository with values. Then, we developed three static web pages to display those issues and their values labels as the prototype of the values dashboard.
In the end, the dashboard prototype contains three views to display the values in issues (\Cref{fig:prototype}):
\begin{enumerate}[leftmargin=0.5cm]
    \item \textbf{Summarised values overview (OV)}.  
    In this view, we display the number of issues of a project that include a given value (e.g., accessibility, pleasure).
    \Cref{fig:prototype-overview} shows this view. This view is commonly provided to provide a quick understanding of the metrics of interest~\cite{Ivanov2019,cauldron,mautic}. 
    We utilised a radar chart to display the number of issues where each value is identified to enable practitioners to compare values that need to be addressed in a project.  
    \item \textbf{Values-labelled list (LI)}. This view (\Cref{fig:prototype-list}) displays a list of issues with their corresponding value labels. We adapted from how GitHub displays issues but with labels or tags specific to human values. We argued that software practitioners are already familiar with this view (\textbf{F3}).  Similar work had used this approach~\cite{Prana2018} that presented the result of automated classification on an artefact from repositories.
    \item \textbf{Values-labelled timeline (TM)}. This view (\Cref{fig:prototype-timeline}) presents the identified values chronologically based on when the issues were posted in the repository. Timeline had been used in previous works ~\cite{Treude2009,Gonz2011} to visualise software evolution. Using this view, we wanted to explore if the emergence order of values during development could benefit software practitioners.
\end{enumerate}

\vspace{-2mm}
\subsection{Data Collection}
\label{subsec:interview}
The first author developed the interview guide and the dashboard prototype in line with the research questions. Then, the interview guide and the dashboard prototype were presented to the other authors for feedback. Afterwards, we made several adjustments based on the feedback. Two authors conducted trial interviews with two software practitioners. Note that we did not include the trial interviews in the data analysis. We also prepared a note to capture ideas from the participants to adapt the questions. We applied some adjustments to the questions and scenarios based on the trial interviews. All authors were agreed on the final version of the interview guide and the prototype. The interview questions are available at \cite{interview-questions}.

Before the interview session started, the participants were requested to read the explanatory statement and fill out the interview's consent form. 
The interviews consisted of three parts. In the first part of the interviews, we asked about the professional backgrounds of the participants. In the second part, we started by describing human values in software development. We also provided some examples of human values considerations in software development. Then, we sought participants' perceptions of human values and whether they considered any human values when developing software (e.g., \textit{based on our examples, do you have any similar experience when developing software?}). In the last part, we explained our proposal of developing a human values dashboard using software development artefacts. More specifically, we first showed some examples of values discussions in GitHub issues. Then, we presented and described our prototype (\Cref{subsec:dashboard}). Finally, we probed for requirements and insights from the participants on how a values-driven dashboard should look like to bring the benefits for software development (e.g., \textit{are there anything you want to have on the dashboard to make it more useful for you?)}

We conducted 15 semi-structured interviews and recorded them using a video conference system with the participants' permission. We did not specify the number of interviews in advance. Instead, we continued recruiting and interviewing in parallel with analysing the data until we reached data saturation~\cite{beitin2012interview, Ournani2020}. We noticed that the convergence of answers and ideas were becoming more apparent in our data analysis after 15 interviews.  
The mean duration time of the interviews is 47 minutes and 7 seconds. Professional transcription services transcribed all audio recordings of the interviews.

\vspace{-2mm}
\subsection{Participants}
\label{subsec:participants}

\subsubsection{Participant Criteria and Recruitment}
Our criteria for recruiting the interview participants were software practitioners who: (1) have been involved in a software development project and (2) are familiar with software development artefacts, including artefacts from software repositories. The rationale for the criteria came from our objective to help practitioners in values integration during software development with a dashboard. 
In particular, we used issues from software repositories as the source of the dashboard.

To recruit the participants, we emailed an invitation to open source project contributors in GitHub. These email addresses were made available by the contributors themselves on their GitHub pages. Interested participants from GitHub were asked to reply to our invitation email. We also published an invitation to our group web page and social media sites, such as LinkedIn and Twitter. Additionally, we asked our colleagues to spread the invitation to their colleague networks. Interested practitioners were asked to inform us of their emails through our colleagues or an online form. We then contacted the participant candidates through email to request consent and arrange the interviews.

\subsubsection{Profile of the Participants}
\Cref{tab:participants} shows the profile of our participants. Our participants had various roles, such as project manager, product owner, system analyst, developer, and user interface (UI) designer. Most of them had less than ten years of experience in software development. Four had more than 15 years of experience. The participants were spread across four continents, but most of them were located in Asia.

\begin{table}[tbh]
\centering
\caption{Profile of the Participants}
\label{tab:participants}
\vspace{-2mm}
{\renewcommand{\arraystretch}{1}
\begin{tabular}{|l|l|r|l|}
\hline
\rowcolor{Gray}
\textbf{ID} & \textbf{Role} & \multicolumn{1}{l|}{\textbf{Experience (year)}} & \textbf{Work Location} \\ \hline
P01 & Developer & 18 & Europe \\ \hline
P02 & System Analyst & 21 & Europe \\ \hline
P03 & Project Manager & 6 & Asia \\ \hline
P04 & System Analyst & 5 & Asia \\ \hline
P05 & Developer & 12 & Australia \\ \hline
P06 & Developer & 3 & Europe \\ \hline
P07 & Developer & 6 & Asia \\ \hline
P08 & UI Designer & 21 & Asia \\ \hline
P09 & Developer & 2 & Asia \\ \hline
P10 & Project Manager & 14 & Asia \\ \hline
P11 & Developer & 3 & Asia \\ \hline
P12 & Developer & 8 & Asia \\ \hline
P13 & Developer & 4 & Australia \\ \hline
P14 & Product Owner & 30 & North America \\ \hline
P15 & Developer & 7 & Asia \\ \hline
\end{tabular}
}
\vspace{-3mm}
\end{table} 

\subsection{Data Analysis}
\label{subsec:analysis}
The thematic analysis approach~\cite{Braun2012} was used to analyse the interview data. The first author conducted a large portion of the analysis. However, the first author consulted with the second and third authors in the case of any doubts or difficulties. Following the thematic analysis approach steps, we started with familiarising ourselves with the interview data by reading the transcriptions and listening to the audio recordings.
Then, we analysed the transcriptions and generated codes and organised them into themes. 
Afterwards, several meetings were organised between the first three authors to review the identified codes/themes and determine the relations between those themes. Finally, we assigned a name and definition for each theme. 
The resulting themes were presented to other authors to obtain feedback. This feedback was then incorporated to create the final themes.

\section{Results}
\label{sec:result}

\subsection{Perceptions towards Human Values}
\label{sec:result-perceptions}
To understand practitioners' perceptions towards human values (\textbf{RQ1}),
we presented 58 human values from Schwartz's model. We observed that the first three participants were overwhelmed by the number of values being presented. Therefore, we decided to only present a set of six values chosen randomly for the remaining participants that include both well-known software engineering (e.g., \textit{privacy}, \textit{security}, \textit{efficiency}) and less-known ones (e.g., \textit{independence}, \textit{wealth}, \textit{sense of belonging}).
Afterwards, we explained some examples of values consideration during software development. 
Then, we probed if the participants have similar experiences in their software development activities. 

The analysis of the interviews indicates that the participants do not understand some values well. The participants also argued that they already considered some of the values during software development. We also found that some values are considered more important than others. We described these findings in the remainder of this section.

\subsubsection{Some values are not well understood by practitioners.}
The participants often asked for an explanation for some values. 
The values that they questioned were the ones that are less known in software engineering, such as \textit{achievements}, \textit{capable}, or \textit{pleasure}. A developer mentioned:
\say{I'm a bit unsure about this area of the achievements and capable means ...}{P01}{Developer}

To explain the values asked by the participants, we used definitions from Schwartz's model ~\cite{Schwartz2012}.
After the explanation, the participants tried to interpret those definitions in software engineering contexts. For example, a developer attempted to relate \textit{wealth} value with development activities:
\say{I would say wealthy it is a bit confusing ... for example, in terms of development activities wealthy can be treated as [paused] so you are writing something, you programmed something, and you were unnecessarily adding something which is going to cost something from the user.}{P05}{Developer}

\subsubsection{Consideration of values during software development.}
The practitioners argued that some human values were already considered during the development of their applications. These values that they have considered were not only the ones that are well-known in software engineering, such as \textit{privacy} or \textit{accessibility}, but also the less-known ones, such as \textit{inclusiveness}, \textit{independence}, and \textit{sense of belonging}. For example, a system analyst mentioned that they developed their project by taking into account the regions, languages, cultures, and educations of their users:
\say{We may develop software for a particular region, a particular crop, maybe a particular season. We integrate all these things during designing ... We consider their language, their education, their culture, their financial capacity, how they can afford.}{P04}{System Analyst}

\subsubsection{Some values are more important than others.} 
The analysis of the interviews shows that the participants are agreed that human values need to be considered in software development. A user interface designer suggested that a software developer should not only focus on the functionality but also on the user of the application:
\say{It shouldn't be only functional, it should be human also, because, at the end of the day, human will use this software, right?}{P08}{UI Designer}

The practitioners also believed that some human values are more important than others depending on the nature of the application being developed. For instance, a UI designer mentioned that if an application is made for the elderly or people with disabilities, then accessibility is more important than the others:
\say{... some project accessibility will be the number one priority like if you say project, with disabled people or aged people, their accessibility should be the number one.}{P08}{UI Designer}
However, in general, the practitioners highlighted that some values are always more important regardless of the application functionality. In our case, those values are the ones that are well-known in software engineering, such as \textit{privacy} and \textit{security}. 

\summary{Our participants deemed human values to be important in software development. However, their understanding of values is limited to those that are well known in software engineering.}

\subsection{Benefits of a Human Values Dashboard}
\label{sec:result-usefulness}
\begin{table*}[htb]
\centering
\caption{Roles and benefit of the dashboard for them 
}
\label{tab:roles_benefit}
\vspace{-2mm}
{\renewcommand{\arraystretch}{1}
\begin{tabular}{|P{4cm}|ll|l|}
\hline
\rowcolor{Gray}
\textbf{Role} & \multicolumn{2}{l|}{\textbf{Benefit}} & \textbf{Soft.Dev. Phase} \\ \hline
\multirow{2}{*}{All Roles} &
\multicolumn{2}{l|}{Align which values are necessary in the project} & Inception \\ \cline{2-4} 
 & \multicolumn{2}{l|}{As a knowledge base for the next stage or future software development} & End of development \\ 
 \hline
\multirow{3}{*}{Project Manager} & \multicolumn{2}{l|}{Support value-based task prioritisation and allocation in project management} & Sprint Planning \\ \cline{2-4} 
 & \multicolumn{2}{l|}{Monitor the  progress of each values-related issue/task} & End of Iteration \\ \cline{2-4} 
 & \multicolumn{2}{l|}{Communicate values with stakeholders} & Release \\ \hline
\multirow{2}{*}{Requirements Engineer} & \multicolumn{2}{l|}{Identify necessary values before starting the implementation} & Requirement \\ \cline{2-4} 
 & \multicolumn{2}{l|}{Ensure identified values administered correctly} & Testing \\ \hline
System Analyst & \multicolumn{2}{l|}{Inform which values are necessary to be taken into  consideration} & Analysis and Design \\ \hline
\multirow{2}{*}{Developer} & \multicolumn{2}{l|}{Support decision in selecting values-specific tasks} & Sprint Planning \\ \cline{2-4} 
 & \multicolumn{2}{l|}{Ensure which values are addressed during development} & Implementation \\ \hline
Tester & \multicolumn{2}{l|}{Ensure all values-related issues/tasks are addressed before release} & Testing \\ \hline
Product Owner & \multicolumn{2}{l|}{Support decision which values are the priority in software} & Inception and Release \\
\hline
General Audience (e.g., in open source   projects) & \multicolumn{2}{l|}{Inform about the focus of values of a software project} & Any phases \\ \hline
\end{tabular}
}
\end{table*}

To understand who will receive the benefits of the dashboard and what the benefits are (\textbf{RQ2}), we demonstrated our dashboard prototype to the participants and asked for their opinions. 
The analysis of the interviews suggested that the identification of human values can generally benefit all roles in software development.
\Cref{tab:roles_benefit} presents the roles and the possible benefits of human values identification for these roles. 
The benefits spanned from the initial stages to the later stages of software development.

In the initial stages of a project's life cycle, being able to detect values potentially helps the whole development team (i.e., \textbf{all roles}) to be aware of which values that need more attention in the project:
\say{So from the beginning, everybody will be in line that on this project, security is our most important concern.}{P08}{UI Designer}
During development (\textit{`mid-flight'}), the identified values in the dashboard can support a \textbf{project manager} to allocate and prioritise tasks based on values:
\say{
Decision on what to do is done by the management. So, they would use this to prioritise what needs to be done first
}{P06}{Developer}
Furthermore, it can potentially help \textbf{requirements engineer}, \textbf{system analyst}, \textbf{developer}, and \textbf{tester} to be aware which values need to be addressed in their respective tasks. Information about values also helps a \textbf{project manager} to communicate with stakeholders (e.g., product owner). The \textbf{product owner} then could make an informed decision about the values priority and the focus of the project:
\say{That is basically what it is up to the program owner to decide the direction the project will take. If he gets all the fire because the application is not stable, probably we want to focus on stability}{P02}{System Analyst}
Even after the project development end, the values information in the software development artefacts is still beneficial as a knowledge base for future software development. For instance, a value identified in an issue could inform \textbf{the development team} how to address similar cases in the future:
 \say{and as I mentioned, these kinds of things also useful for learning for future projects.}{P09}{Developer}

We also discovered the dashboard's benefits to open source projects. Because their development artefacts are publicly available, \textbf{general audience} who are interested in an open-source project can use the dashboard to determine which values become the focus of the project. For example, a \textbf{user} who wants to be a \textbf{contributor} could figure out the extent to a project cares about a particular value:
\say{... and then when it comes to a person who is outside of a repo, ... they will also get an insight on this. You like, this is very informative and from a perspective of a person who wants a clear summarisation on what is happening inside a repository, and what are the areas that it is being the main focus}{P07}{Developer}

\summary{A values dashboard is considered useful for various roles in various phases in software development. It can be used especially to determine values-driven priorities in a project and raise awareness of the development team about values.}

\subsection{Artefacts for the Dashboard Source}
\label{sec:result-artefacts}
To answer \textbf{RQ3}, which artefacts are suitable for the dashboard, we presented some examples of how values are being discussed in a GitHub issue. We then probed the participants with other artefacts that are generated during software development. 
Afterwards, we asked which artefacts that they think are suitable for the identification of values. 

\Cref{tab:artefacts-simple} shows software development artefacts that the participants think are suitable as a source for values identification. 
These artefacts are generated in different phases of software development. 
We found that these artefacts were not necessarily created in every software project. It depends on how the development team manages the project. For instance, an open-source project may not consider market research documents as necessary. On the other hand, some artefacts, such as issue discussions, are generated when the development team contributes to the software repository.

\begin{table}[htb]
\vspace{-3mm}
\centering
\caption{Potential Artefacts for Human Values Identification
\newline \small{\textbf{*} denotes more suitable}
}
\label{tab:artefacts-simple}
\vspace{-2mm}
{\renewcommand{\arraystretch}{1}
\begin{tabular}{|l|l|}
\hline
\rowcolor{Gray}
\textbf{Artefact} & \textbf{Soft.Dev. Phase} \\ \hline
Market research documents & Beginning of the project \\ \hline
Requirements documents* & Requirements \\ \hline
Design documents & Analysis and Design \\ \hline
Features specification documents & Analysis and Design \\ \hline
Issue discussions* & Implementation \\ \hline
Pull request discussions & Implementation \\ \hline
\end{tabular}
}
\vspace{-3mm}
\end{table}

\subsubsection{Reasons for selection}
We found that the participants chose the artefacts in \cref{tab:artefacts-simple} for two main reasons. \textbf{\textit{Firstly, values can be identified within the artefacts}}. For example, a UI designer believed that \textbf{market research documents} that are written before the application development should capture the values of the users:
\say{So it will be in the research report, and that will honour the value of the user.}{P08}{UI Designer}
Other artefacts, such as \textbf{pull requests}, have a discussion functionality that captures the conversation between stakeholders. A developer mentioned that it is possible to identify values in these conversations as well:
\say{So pull requests are the areas that we are having most of the conversations as well. ... Like if we are having pull requests with security issues all the time.}{P07}{Developer}

The \textbf{\textit{second reason}} the participants chose those artefacts because \textbf{\textit{the artefacts are used and referred during software development}}. The participants agreed that identifying values in these artefacts would support them in addressing values in software development. For example, a developer mentioned that the identification of values in \textbf{issue discussions} would benefit them:
\say{[repository] is the place where most developers, especially us, busy with almost the day, every day. So, if we see this kind of issue tracking, that kind of thing happening in the [repository] site, I think it would be, especially for developers, it would be good.}{P05}{Developer}
Another practitioner also mentioned that the identification of values in \textbf{design documents} could assist in planning the software development:
\say{So at the design level, we can target those areas where the accessibility issues can be pointed out so that the things are more planned accordingly in terms of accessibility.}{P10}{Project Manager}

\subsubsection{Suitability for the identification of values}
We also asked the practitioners if some artefacts are \textbf{\textit{more suitable}} as the target of values identification. Some practitioners argued that \textbf{requirements documents} is more suitable. The reason for choosing this artefact was that these documents are created at the initial stage of development, such that the identification of values can inform the next development stages as well:
\say{I think it will be better to run this on requirements because then we can identify everything at the initial stage.}{P13}{Developer}
However, a developer reminded that a bias might exist in the requirements because it was usually developed by only either one or a few persons:
\say{... like for the requirements and stuff, it will be most probably one person who is getting involved ... that will be pretty biased.}{P07}{Developer}

Another more suitable artefact is \textbf{issue discussions}. A reason for this opinion was that it is the place where stakeholders and contributors discuss different aspects of an application:
\say{Actually, I would say it would be issues and the public's comments. ... those are the basis that people engage from different aspects.}{P07}{Developer}
Furthermore, the identification of values in this artefact will help developers identify the impacts of an issue that needs to be addressed:
\say{... until that point, whatever we did, is there any sort of violation, is there any sort of lacking that we may be unconsciously generated or produced? So, we can take that from there. .. we can concentrate, we can fix that before the release, for example}{P05}{Developer}

\summary{A wide range of software development artefacts can be used as a data source for a values dashboard. However, our participants particularly mentioned that requirements documents and issue discussions are the most suitable sources.}

\subsection{What is Needed in a Values Dashboard}
\label{sec:result-requirements}

To answer \textbf{RQ4}, what is needed for a human values dashboard to be helpful in software development, the prototype was used as a trigger for a discussion. 
In the discussion, we probed what is required for a values dashboard to support various roles in conducting their software development activities.

\begin{table}[htb]
\centering
\caption{Proposed requirements for the dashboard.}
\label{tab:requirements}
\vspace{-2mm}
{\renewcommand{\arraystretch}{1}
\begin{tabular}{|l|P{7.5cm}|}
\hline
\multicolumn{2}{|l|}{\cellcolor{Gray}Requirements} \\ \hline
R1 & The identification of values in the dashboard shall be conducted   automatically \\ \hline
R2 & It should maintain the traceability between the identified values and their artefact source \\ \hline
R3 & It shall allow the development team to determine the values priority of a project \\ \hline
R4 & It shall display the artefacts based on the values priority determined in   a project \\ \hline
R5 & It shall inform the latest update on the artefacts where values are identified in a project \\ \hline
R6 & It shall provide different views for various roles to support addressing values in software development \\ \hline
\end{tabular}
}
\vspace{-3mm}
\end{table}

\Cref{tab:requirements} shows what our practitioners believed is required for the dashboard. We formulated their suggestions as requirements by using imperative words \cite{hammer1998doing}. 
For the first requirement (\textbf{R1}), our participants believed that it would be helpful if the identification of values is conducted automatically. This functionality will reduce their effort on manual identification and provide third-party opinion on the work: 
\say{I know in [repository], they also have worked with, try to tag issues and categorize them manually, but that often either is forgotten or it somehow fails.
}{P01}{Developer}
Surprisingly, although the participants were aware of the limitations of an automated approach, they were still interested in using it:
\say{Also, even though machine learning is not always correct, you might also find that it sees things that you do not.}{P01}{Developer}

The second requirement (\textbf{R2}) were asked by our participants to navigate to the artefacts where values are identified: 
\say{If we have a report mentioning all the details, at least the file where it is happening right now would be good for us.}{P05}{Developer}
This functionality will be particularly helpful if the dashboard and the artefacts are in different systems.

Our participants believed that some values are more important than others. Therefore, they requested functionality to specify which values they consider important for their project (\textbf{R3}). This functionality will help them to support value-driven decision making, such as prioritising which values that they want to address:
\say{As I said, I want to prioritise security number one, visibility number two, pleasure number three, wealth number four. This is very important.}{P08}{UI Designer}

In \textbf{R4}, our participants wanted the dashboard to display the artefacts based on the priority of the values determined for a project. This functionality will help them to focus on specific values and reduce information overload:
\say{But as I said before, I like the idea of adding the feature that you could pick, which tag [of values] you are interested in and then only see them.}{P06}{Developer}
A values dashboard must also inform the development team if there is a change to an artefact where values are identified (\textbf{R5}). This information will keep the team up to date with the current status of values identified throughout the application development. 
This functionality can be implemented by displaying the last activity or sending a notification when there is an update on an artefact:
\say{I mean, you could also add a notification. It is just my guess, I mean, if it is a security thing that has been notified somewhere, or somebody wrote something that flags a security issue, you would want to know, put that right away}{P06}{Developer}

Our participants agreed that a values dashboard should cater to various roles in software development (\textbf{R6}). 
The multiple views that we provide in the prototype (\Cref{fig:prototype}) can support these roles in addressing values during software development.
For example, the summarised values overview (\textbf{OV}) provided in the prototype (\Cref{fig:prototype-overview}) informs a project manager about the state of values identified in a project before going into the details:
\say{As a team leader or a project manager, I would prefer the second one [\textbf{OV}] because I will be going through the tickets all the time ... So I will just get an overview of what I am working on.}{P07}{Developer}
The list view or \textbf{LI} (\Cref{fig:prototype-list}) will support system analysts, developers, and testers to ensure values are incorporated in the project:
\say{And after that, they will also use this information [\textbf{LI}] for their QA [Quality Assurance]. And they said, okay, we have to test this thing. Okay. Is this a security issue? Is accessibility issue or anything else? They can use all this information.}{P11}{Developer}
Testers can use the timeline view or \textbf{TM} (\Cref{fig:prototype-timeline}) to monitor the artefact where values are identified before conducting their tasks:
\say{For a tester, I would say a timeline is best ... they can use the timeline to see what issues are currently in progress and what are closed and ready to test.}{P13}{Developer}

The first two requirements (R1 and R2) support the development team to identify values and their corresponding artefacts efficiently.
The following two requirements (R3 and R4) utilise values as prioritisation criteria in the development activities. Meanwhile, the remaining requirements, R5 and R6, are in line with the benefit of a dashboard to promote the awareness of values among software development teams. 

\summary{
Six high-level requirements are suggested to be necessary for a values dashboard. These requirements are considered to support various roles to address values during software development.}
\section{Discussion}
\label{sec:discussion}

\subsection{The Awareness of Values}
Our analysis showed that software practitioners are familiar with only a limited set of values, such as \textit{security} and \textit{accessibility}. 
This finding strengthens the findings from \cite{Perera2020}, which found that only a few values were discussed in recent academic software engineering papers. 
One possible reason for this stems from the fact that there is a lack of understanding of these values in the software engineering context. Additionally, our participants also thought that the values that they are familiar with are important and believed that they already consider these values during software development. The other values that our participants are not too familiar with become a `nice to have' in an application. These findings show the need to increase practitioners' awareness of human values. A possible solution is to provide a contextualised software engineering definition for each of these values \cite{Mougouei2018,Perera2020}. Additionally, a tool, such as a values dashboard discussed in this study, can be utilised as it was found in this study to have the benefit of increasing awareness (e.g., of values) among the development team \cite{Treude2009}.

\subsection{A Values Dashboard for Users}
Our results suggested that one of a human values dashboard's main benefits is to promote the awareness of values. This awareness of values could trigger discussions among stakeholders on what values must be considered in an application.
Then, the development team can focus on the prioritised values and ensure these values are addressed during development.
Nevertheless, this study was focused on software practitioners who are involved in software development. 
This study did not include users as one of the stakeholders in software development yet. 
Users can potentially have access to a human values dashboard (e.g., in open source projects). 
In this case, a future study can be carried out first to collect their needs and then investigate the benefits of a human values dashboard to users. For example, whether a human values dashboard can help users evaluate the values of an application \cite{Kujala2009} and choose their preferred application \cite{Wang2013,Harris2016,Fu2013}.

\subsection{Artefacts as the Source for the Dashboard}
Software practitioners in our study suggested that requirements documents and issue discussions are considered more suitable for mining values. 
In this case, we believe that issue discussions have several advantages over other artefacts. For instance, it is common for developers to have source code changes linked to an issue in a repository \cite{Bissyande2013}. This practice allows them to obtain detailed information on the issue that they need to address. Such capabilities were requested by the interviewees (\textbf{R2} in \Cref{tab:requirements}). On the other hand, requirements documents tend to be independent such that tracing requirements is a challenge in software engineering research \cite{Kasauli2017}. 
An advantage for open source projects, opening the issue discussions publicly allows users to provide feedback. 
Users' feedback is a potential place where values can be found \cite{Shams2020}.
Therefore, we recommend that future research compares the extent to which human values can be identified in issue discussions and other artefacts.

\subsection{Implementing a Values Dashboard}
We figure that the requirements suggested by our participants are high-level requirements and could be implemented in various ways. In this case, our prototype can be used as a starting point for implementing the front end of a human values dashboard. 
For the back end, we believe further studies are necessary to develop an approach to automatically identify values (\textbf{R1} in \Cref{tab:requirements}). 
For a project with an existing dashboard, we believe that it is possible to implement a human values dashboard to complement the existing one. Functionalities of such dashboards, e.g., progress monitoring \cite{cauldron,Ivanov2019}, could benefit from the perspective of the values offered by our visioned values dashboard. For instance, the development team can monitor the progress of values-related tasks.
Thus, future work is recommended to implement the values dashboard or integrate the values dashboard into an existing software development dashboard. 

\subsection{Limitations of the Dashboard}
Our visioned dashboard in \Cref{fig:proposed-dashboard} may have several limitations.
Firstly, the dashboard depends on the availability of the artefacts. A project may not have all the artefacts in \Cref{tab:artefacts-simple} subject to how the project is managed. However, we believe that this dashboard will be potentially useful for projects with substantial numbers of software development artefacts. Secondly, the benefit of the dashboard will depend on how values are identified in software development artefacts. Manual identification of values will need considerable effort, which is not favourable for the development team. Meanwhile, identification using an automated approach has limitations in accuracy (\Cref{sec:result-requirements}).
Although this limitation is understandable by the practitioners, further study is necessary to understand how this will affect the use of the dashboard. 
\vspace{2mm}
\section{Threats to Validity}
\label{sec:threats}

We discussed the potential threats arising from our research method and the findings using the following validation criteria that are considered more suitable for qualitative research \cite{Guba1981, Stol2014, Cruzes2011}.

\textbf{Credibility}:
The possible threats to the credibility of this study could arise from the procedure used for collecting the data, developing the interview questions, and selecting the participants. While we only collected data from one source (the interviews), we believe our initial step of examining the literature before developing the dashboard prototype could increase the plausibility of our findings.
To mitigate the threats coming from the interview questions, we used open-ended questions and asked follow-up questions tailored to each participant's responses. 
To reduce possible threats from the selection of participants, we relied on our criteria for the participants' recruitment. Our list of participants consists of software practitioners with diverse roles, experiences, and work locations. Thus, we believe that our participants had the right competencies to provide insights into our study.
To mitigate the uneven number of participants in each role, we also asked the participants to share their opinions from other roles' perspectives. This approach also allows cross-validation of findings across different roles. 

\textbf{Confirmability}: 
A possible threat to the confirmability of this study might be introduced by the definition of human values, which are not specifically developed for software engineering. To mitigate this, we provided examples to the participants to describe what a value possibly means in software engineering contexts (e.g., `\textit{user who values privacy may not choose an application with a bad privacy reputation}'). 
We also allowed participants to reflect and translate values into contextualised software engineering definitions based on their experiences. 
The data analysis could introduce another possible threat to confirmability as the first author primarily conducted it. We mitigate this threat by having other authors reviewed and validated the codes/themes in several discussions.

\textbf{Transferability}:
We accept that our findings cannot be generalised to all software organisations and practitioners. Different results might be discovered if we have included another group of participants. However, we reduced this threat by involving a reasonable number of participants with various development roles and work locations. Additionally, we also observed the convergence of our data during the interviews and data analysis. We cannot also generalise the relative importance of a specific value against the others because we only presented a small set of values to the participants. However, we still found that some values are more important than others.
\vspace{-2mm}
\section{Conclusion}
\label{sec:conclusion}
In this work, we visioned a values-driven dashboard and investigate whether it can be helpful to support software practitioners to handle values during software development. 
We found that, in general, the practitioners acknowledged that a values dashboard would be beneficial for them. The dashboard could raise awareness of values among the development team and inform values-based decision making in project management. 
Supporting our idea of using artefacts as the dashboard source, practitioners suggested requirement documents and issue discussion as the most suitable for values identification in the dashboard.
We also received suggestions as a set of requirements to develop our visioned dashboard. 

The results of this study could be used for the development of a human values dashboard. Future works will be focused on developing a human values dashboard and investigating issue discussions in software repositories as the data source for the dashboard. Other studies are also necessary to investigate the possibility of using automated approaches for the values identification. 

\bibliographystyle{ACM-Reference-Format}
\bibliography{ms}

\end{document}